\documentclass[prb,preprint,superscriptaddress,groupedaddress]{revtex4}
\usepackage{graphicx}
\usepackage{dcolumn}
\usepackage{bm}
\usepackage{comment}
\usepackage{color}
\usepackage{amsmath}
\usepackage{accents}
\usepackage[utf8]{inputenc}
\usepackage{subfigure}

\newcommand{\be}{\begin{equation}}
\newcommand{\ee}{\end{equation}}
\newcommand{\bea}{\begin{eqnarray}}
\newcommand{\eea}{\end{eqnarray}}
\newcommand{\acosh}{\mathrm{acosh}}

\newcommand{\sn}{\mathrm{sn}}
\newcommand{\cn}{\mathrm{cn}}
\newcommand{\dn}{\mathrm{dn}}

\begin{document}
\title{Spin Torque Oscillators with Thermal Noise: A Constant Energy Orbit Approach}
\author{D.~Pinna}
\email{daniele.pinna@nyu.edu}
\affiliation{Department of Physics, New York University, New York, NY 10003, USA}
\author{D.~L.~Stein}
\affiliation{Department of Physics, New York University, New York, NY 10003, USA}
\affiliation{Courant Institute of Mathematical Sciences, New York University, New
             York, NY 10012, USA}
\author{A.~D.~Kent}
\affiliation{Department of Physics, New York University, New York, NY 10003, USA}

\begin{abstract}
We study the magnetization dynamics of spin torque oscillators in the presence of thermal noise and as a function of the spin-polarization angle in a macrospin model. The macrospin has biaxial magnetic anisotropy, typical of thin film magnetic elements, with an easy axis in the film plane and a hard axis out of the plane. Using a method that averages the energy over precessional orbits, we derive analytic expressions for the current that generates and sustains out-of-plane precessional states. We find that there is a critical angle of the spin-polarization necessary for the occurrence of such states and predict a hysteretic response to applied current. This model can be tested in experiments on orthogonal spin-transfer devices, which consist of both an in-plane and out-of-plane magnetized spin-polarizers, effectively leading to an angle between the easy and spin-polarization axes.
\end{abstract}

\pacs{Valid PACS appear here}
\maketitle


\section{Introduction}

Magnetic excitations in spin valves and magnetic tunnel junctions present a
set of phenomena that are of considerable interest, both for the
new physics they present and for a variety of potential
applications. More specifically, the effects of thermal noise, combined with spin 
transfer torques induced by a current, present novel phenomena.
On the technological side, large-angle steady-state magnetic
excitations in spin-valves and magnetic tunnel junctions induced by dc~currents
has recently attracted much attention~\cite{Krivorotov,Houssameddine}. In
conjunction with their magneto-resistance~(MR) response, persistent
magnetization oscillations could  lead to wide-band tunable RF oscillators~\cite{SlonPatent} operating in the
GHz to THz frequency range. To these ends, it is of importance to understand the physics of
current induced magnetic excitations in the presence of noise and to understanding 
the factors that determine the tunability and quality factors of
these systems.

Within a macrospin picture, current-induced steady-state motion appears when the magnetization settles into a stable oscillatory trajectory that balances the spin-torque and damping~\cite{Bertotti05}. The oscillatory behavior is magnetization precession at a frequency associated with the element's magnetic anisotropy, which can arise, for example, because of the element's shape (i.e. magnetic shape anisotropy) or magnetocrystalline anisotropies. Thermal noise can, however, can alter the frequency and amplitude of the motion as well as change the conditions under which steady-state precession occurs. As a result, it is important to know both how an applied current will influence the amplitude and frequency of a stable magnetic oscillation and how thermal noise will perturb this configuration by inducing amplitude and phase noise.

If amplitude and phase diffusion due to spin-torque and thermal noise effects occurs on a timescale much larger than that of magnetization precession, it becomes possible to analyze the steady-state dynamics perturbatively~\cite{JooVonPRL}.  In this case, the magnetization dynamics will consist of a fast gyromagnetic precession whose amplitude slowly changes over time due to spin-torque and thermal effects. This has successfully been used to study the dynamical and thermal stability of nanomagnets subject to spin-polarized currents~\cite{STIEEE08}. This separation of dynamical timescales falls under the framework of multiscale analysis, which can be applied in various ways. 

Three different approaches have been proposed in the literature in the context of spin-transfer. Apalkov and Visscher~\cite{Apalkov} employed an effective Fokker-Planck (FP) equation, which described the diffusion of a macrospin's energy under the influence of both spin-transfer torque and thermal noise. This has been used to interpret results on studies of thermally activated magnetic switching~\cite{Bedau,TaniguchiE}. Kim, Slavin and Tiberkevich~\cite{JooVonPRL,STIEEE} studied the Landau-Lifshitz-Gilbert-Slonczewski (LLGS) equation by noting its analogy to the van der Pol oscillator equation~\cite{Strogatz}. This resulted in an elegant treatment of the leading nonlinear effects governing the oscillatory equilibrium steady-state dynamics of the spin-wave eigenmodes. The approach~\cite{Exp3} has had success in explaining the experimentally observed dependence of the oscillator's output power on bias current for spin-valves and magnetic tunnel junctions~\cite{Exp1,Exp2,Exp4,Exp5,Exp6}, as well as providing a framework for the extension of multiscaling methods to spatially extended magnetic systems in which multiple coupled spin-wave modes may be excited~\cite{Aguiar}.

Finally, macrospin dynamics subject to thermal noise have been modeled using a stochastic Langevin equation for the time evolution of the macrospin energy by Newhall and Vanden-Eijnden~\cite{Katie} and in previous work by the Authors~\cite{IEEE}. This reduces the complexity of the LLGS equations to a 1D stochastic differential equation. Stochastic energy space dynamics have been used to describe the full nonlinear dependence of mean switching time on applied current~\cite{PRB} for biaxial macrospin models ($\log\tau\propto(1-I)^{\beta(I)}$) as an analytic continuation of the uniaxial macrospin model. Recently, Dunn and Kamenev have extended this approach to propose AC current-driven resonant switching~\cite{Dunn}. 

Recent research on spin-torque oscillators has focused on the excitation of stable in-plane (IP) and out-of-plane (OOP) precession about the easy and hard magnetic anisotropy axes of thin film nanomagnets with biaxial magnetic anisotropy. In this Article we present a stochastic theory of these precessional dynamics valid over a wide range of parameters. We focus on the OPP dynamics and show the conditions under which precessional motion about the hard axis occurs. The oscillator behavior we find is reminiscent of that observed in experiments on a spin-valve where spin-torque effects are due to the influence of both a perpendicularly magnetized polarizer and in-plane magnetized reference layer~\cite{Houssameddine}. The two contributions lead to a net spin-torque which can be formally thought to arise from a tilted spin polarizer\cite{ChinaTilt1,ChinaTilt2,ChinaTilt3,ChinaTilt4}. The precessional dynamics are found to be stable at room temperature and, as a result, have great potential for the development of spin-torque nano-oscillators.  


\section{General Formalism}

We study a monodomain of magnetization $\mathbf{M}$ of constant modulus ($M_S=|\mathbf{M}|$) with a biaxial magnetic anisotropy, with easy direction $\mathbf{\hat{n}}_{K}$ and hard direction $\mathbf{\hat{n}}_{D}$. Its energy landscape depends on the projection of the magnetization onto these two axes. We write the easy and hard axis anisotropy energies as $K=(1/2)\mu_0 M_SH_KV$ and $K_M=\mu_0 M_S^2V$, where $H_K$ is the anisotropy field and $V$ is the volume of the magnetic element. To lowest order, in the absence of external magnetic fields and magnetic dipole fields arising from other magnetic layers, the energy can be written as:
\begin{equation}
U(\mathbf{m})=K\left[D(\mathbf{\hat{n}}_D\cdot\mathbf{m})^2-(\mathbf{\hat{n}}_K\cdot\mathbf{m})^2\right],
\end{equation}
where $\mathbf{m}=\mathbf{M}/|\mathbf{M}|$ is the normalized magnetization vector and $D\equiv K_M/K=M_S/H_K$ is a dimensionless ratio of the two anisotropy constants. This energy has minima and thus stable magnetic configurations for $\mathbf{m}$ parallel and antiparallel to $\mathbf{\hat{n}}_K$.

The evolution of such a macrospin subject to thermal noise and spin-transfer torques is described by
a stochastic Landau-Lifshitz-Gilbert-Slonczewski (LLGS) equation of the form
\begin{equation}
\label{eq:Langevindynamics}
\dot{m}_i=A_i(\mathbf{m})+B_{ik}(\mathbf{m})\circ H_{th,k}
\end{equation}
where the stochastic contribution $\mathbf{H}_{th}$ is taken to have zero mean and delta-function correlation $\langle H_{th,i}(t)H_{th,k}(t')\rangle=2C\delta_{i,k}\delta(t-t')$. The diffusion constant $C=\frac{\alpha}{2(1+\alpha^2)\xi}$ (with $\xi\equiv K/k_BT$ the energy barrier height divided by the thermal energy) is chosen to satify the fluctuation-dissipation theorem, and multiplicative noise `$\circ H_{th,k}$' is interpreted in the Stratonovich sense~\cite{Karatsas}. The expressions for the drift vector $\mathbf{A}(\mathbf{m})$ and diffusion matrix $\hat{\mathbf{B}}(\mathbf{m})$ terms, written in vectorial form, read:

\begin{eqnarray}
\mathbf{A}(\mathbf{m})&=&\mathbf{m}\times\mathbf{h}_{\mathrm{eff}}-\alpha\mathbf{m}\times\left(\mathbf{m}\times\mathbf{h}_{\mathrm{eff}}\right)\nonumber\\
&-&\alpha I\mathbf{m}\times\left(\mathbf{m}\times\mathbf{\hat{n}}_p\right)-\alpha^2 I\mathbf{m}\times\mathbf{\hat{n}}_p,\nonumber\\
B_{ik}(\mathbf{m})&=&\sqrt{\frac{\alpha}{2\xi(1+\alpha^2)}}[-\epsilon_{ijk}m_j-\alpha(m_i m_k - \delta_{ik})].
\end{eqnarray}
where $\mathbf{h}_{\mathrm{eff}}=-\frac{1}{\mu_0 M_S H_K V}\nabla_{\mathbf{m}}U(\mathbf{m})$ is the effective field rescaled by $H_K$, $I=q(\hbar/2e)\eta J/(\alpha \mu_0 M_S H_K d)$, with $d$ the thickness of the magnetic free layer, is a natural current scaling with $\eta = (J_{\uparrow}-J_{\downarrow})/(J_{\uparrow}+J_{\downarrow})$, the
spin polarization of incident current density $J$ along polarization axis $\mathbf{\hat{n}}_p$ and $q$ a normalization constant which will be discussed below. The temporal derivatives appearing in (2) and throughout this paper are with respect to the natural timescale $\tau=(\gamma/(1+\alpha^2))\mu_0 H_K t$, where $\gamma$ is the gyromagnetic ratio. The dynamics associated with (\ref{eq:Langevindynamics}) \cite{Palacios,LiZhang,Apalkov} result in Boltzmann equilibrium conditions at long times.  
When $\mathbf{\hat{n}}_K$ and $\mathbf{\hat{n}}_D$ lie perpendicular to each other (such as in typical spin-valves), the macrospin's geometry is fully determined by two angles: $\omega$ the angle between the spin-polarization axis $\mathbf{\hat{n}}_p$ and $\mathbf{\hat{n}}_K$ and
the azimuthal angle $\psi$ characterizing the extent to which $\mathbf{\hat{n}}_p$, $\mathbf{\hat{n}}_K$ and $\mathbf{\hat{n}}_D$ are coplanar~(see Fig. 1). We choose a coordinate frame where $\mathbf{\hat{n}}_K$ and $\mathbf{\hat{n}}_D$ define the x- and z- axes respectively.

A tilted spin-polarization axis allows modeling a spin-torque that results from more than one ``polarizing'' layer in a spin-valve (or MTJ) stack or, more generally, a free layer that has an easy-axis tilted relative to the spin-polarization axis. This is particularly relevant to experiments employing a perpendicular polarizer layer with an in-plane magnetized spin-valve, consisting of a free and reference layer~\cite{Ozy, Ebels, Firastrau, Papusoi, Liu1, Liu2, Ye}. In this case, the effective spin-polarization will be tilted with respect to the easy-axis of the free layer. The net spin polarization axis can be written as:
\be
\mathbf{\hat{n}}_p=\frac{\eta_\mathrm{ref}\mathbf{\hat{n}}_\mathrm{ref}+\eta_\mathrm{pol}\mathbf{\hat{n}}_\mathrm{pol}}{\sqrt{\eta^2_\mathrm{ref}+\eta^2_\mathrm{pol}}},
\ee 
where $\mathbf{\hat{n}}_\mathrm{ref}$ and $\mathbf{\hat{n}}_\mathrm{pol}$ are the spin-polarization axes directions of the reference and polarizer layers. The tilt angle $\omega$ can then be written in terms of the ratio of the spin-torque efficiencies $\omega=\mathrm{atan}(\eta_\mathrm{pol}/\eta_\mathrm{ref})$. The normalization factor $q=\sqrt{\eta^2_\mathrm{ref}+\eta^2_\mathrm{pol}}$ appears in the definition of the applied current $I$ discussed earlier.
\begin{figure}
	\begin{center}
	\centerline{\includegraphics [clip,width=4in, angle=0]{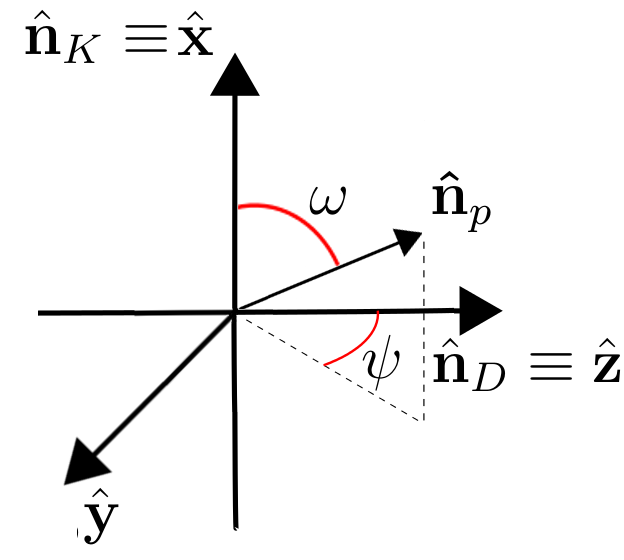}}
	\caption{Uniaxial easy $\hat{n}_K$ and hard-axis $\hat{n}_D$ magnetic anisotropy directions are shown along with spin-polarization direction $\hat{n}_p$. 
	The spin-polariation is tilted by an angle $\omega$ with respect the magnetic easy axis.}
	\label{F1}
	\end{center}
\end{figure}

All numerical results we present have been obtained by solving~(\ref{eq:Langevindynamics}) 
for ensembles of 5120 independent macrospins using an integration time step of $0.01$ in natural time, i.e. $\tau$. For concreteness, we set 
the damping constant $\alpha=0.04$ and barrier height $\xi=80$.


\section{Energy-Averaged Dynamics}

In the absence of damping and thermal noise, the dynamics~(\ref{eq:Langevindynamics})
preserve the macrospin's energy which, expressed in dimensionless form, reads:	

\be
\label{eq:epsilon1}
\epsilon=\frac{U(\mathbf{m})}{K}=Dm_z^2-m_x^2,
\ee
The conservative trajectories come in two different types. For $-1<\epsilon<0$ the magnetization gyrates
around the easy axis $\mathbf{\hat{n}}_K$ and is said to be precessing ``in-plane" (IP). For $0<\epsilon<D$, the magnetization precesses about the hard axis $\mathbf{\hat{n}}_D$ and is said to be precessing ``out-of-plane" (OOP).
The evolution of such trajectories can be described analytically by solving the LLGS equation in the absence 
of noise, damping and spin-transfer torque:~\cite{Serpico}

\bea
\dot{m}_x^0&=&-Dm_z^0m_y^0\nonumber\\
\dot{m}_y^0&=&(D+1)m_z^0m_x^0\nonumber\\
\dot{m}_z^0&=&-m_y^0m_x^0
\eea

For IP trajectories~\cite{PRB} one has

\begin{eqnarray}
m_x^0(t)&=&\pm \sqrt{\frac{D-\epsilon}{D+1}}\;\dn\left[\sqrt{D-\epsilon}t,k_\mathrm{IP}^2\right]\\
m_y^0(t)&=&\sqrt{1+\epsilon}\;\sn\left[\sqrt{D-\epsilon}t,k_\mathrm{IP}^2\right]\\
m_z^0(t)&=&\sqrt{\frac{1+\epsilon}{D+1}}\;\cn\left[\sqrt{D-\epsilon}t,k_\mathrm{IP}^2\right],
\end{eqnarray}
where $k_\mathrm{IP}^2\equiv D\frac{1+\epsilon}{D-\epsilon}$ and $\sn[\cdot],\dn[\cdot],\cn[\cdot]$ are Jacobi elliptic functions~\cite{Abramowitz}. The period of these trajectories as a function of energy can be expressed as a complete
elliptic integral of the first kind:
\be
\label{eq:period_Tmin}
T(\epsilon)=\frac{4}{\sqrt{D-\epsilon}}\int_0^1\frac{dx}{\sqrt{(1-x^2)(1-k_\mathrm{IP}^2x^2)}}=\frac{4}{\sqrt{D-\epsilon}}\mathrm{K}(k_\mathrm{IP}^2).
\ee
The amplitudes of an orbit's precession, projected onto the $\hat{\mathbf{z}}$-$\hat{\mathbf{y}}$ plane, are\footnote{precession is around the $\hat{\mathbf{x}}$-axis}

\bea
A_{\hat{\mathbf{z}}}(\epsilon)&=&\sqrt{\frac{1+\epsilon}{D+1}}\\
A_{\hat{\mathbf{y}}}(\epsilon)&=&\sqrt{1+\epsilon}
\eea
Analogously, for OOP trajectories

\begin{eqnarray}
\label{eq:posstates}
m_x^0(t)&=&\sqrt{\frac{D-\epsilon}{D+1}}\;\cn\left[\sqrt{D(1+\epsilon)}t,k_\mathrm{OOP}^2\right]\\
m_y^0(t)&=&\sqrt{\frac{D-\epsilon}{D}}\;\sn\left[\sqrt{D(1+\epsilon)}t,k_\mathrm{OOP}^2\right]\\
m_z^0(t)&=&\pm \sqrt{\frac{1+\epsilon}{D+1}}\;\dn\left[\sqrt{D(1+\epsilon)}t,k_\mathrm{OOP}^2\right],
\end{eqnarray}
with $k_\mathrm{OOP}^2\equiv \frac{D-\epsilon}{D(1+\epsilon)}$. Period and projected precession amplitudes in the $\hat{\mathbf{x}}$-$\hat{\mathbf{y}}$ plane are:
\bea
\label{eq:period_Tplus}
T(\epsilon)&=&\frac{4}{\sqrt{D(1+\epsilon)}}\int_0^1\frac{dx}{\sqrt{(1-x^2)(1-k_\mathrm{OOP}^2x^2)}}=\frac{4}{\sqrt{D(1+\epsilon)}}\mathrm{K}(k_\mathrm{OOP}^2)\\
A_{\hat{\mathbf{y}}}(\epsilon)&=&\sqrt{\frac{D-\epsilon}{D}}\\
\label{eq:Zamplitude}
A_{\hat{\mathbf{x}}}(\epsilon)&=&\sqrt{\frac{D-\epsilon}{D+1}}.
\eea

A sample of these trajectories for positive and negative energies is shown in Fig.~2, and orbital frequency as a function of energy is plotted in Fig.~3. 
The unit magnetic sphere can be separated into four distinct basins, two corresponding
to $\epsilon<0$ dynamics and the others two to $\epsilon>0$. For large values of $D$ the $\epsilon>0$ OPP basin can lead to a larger oscillatory
resistance signals than $\epsilon<0$ IP basin due to the larger precessional amplitudes ~(\ref{eq:Zamplitude}).

\begin{figure}
	\begin{center}
	\centerline{\includegraphics [clip,width=4in, angle=0]{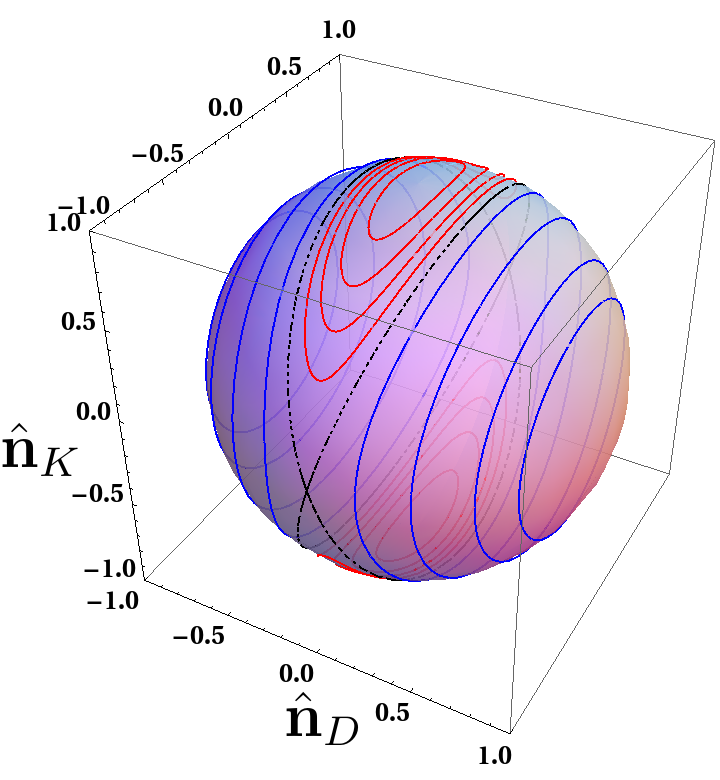}}
	\caption{Constant energy trajectories for $D=10$. $\epsilon<0$ trajectories are shown in red whereas $\epsilon>0$ trajectories are shown in blue. Notice how two distinct basins exist for positive and negative energy trajectories. The singular separatrix, corresponding to $\epsilon=0$, separating the different basins is shown in black.}
	\label{F2}
	\end{center}
\end{figure}

\begin{figure}
	\begin{center}
	\centerline{\includegraphics [clip,width=4in, angle=0]{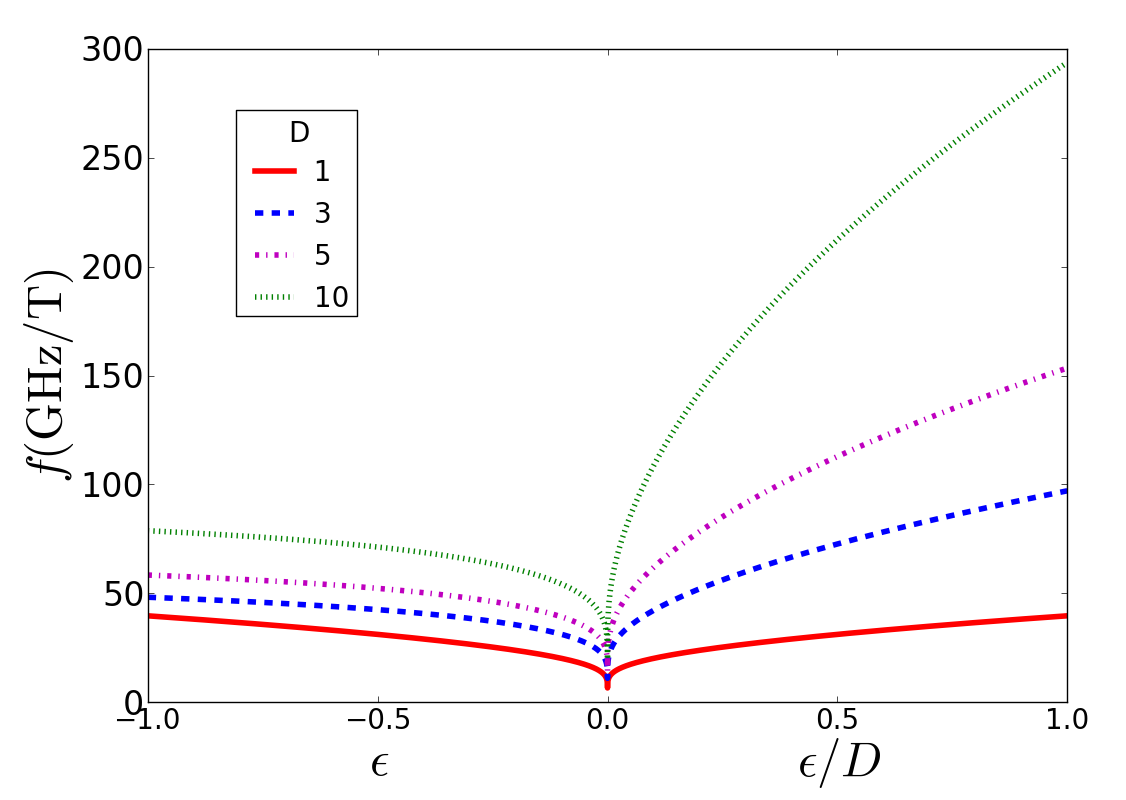}}
	\caption{Orbital frequencies plotted as a function of $\epsilon$ for different $D$. To compare the results, the positive portion of $\epsilon$ axis has been rescaled by $D$. Frequency is expressed in units of (GHz/T). Physical frequency is obtained upon multiplying by $\mu_0 H_K$. The sharp minimum in the frequency is a result of the precessional period diverging at $\epsilon=0$.}
	\label{F3}
	\end{center}
\end{figure}

Upon introducing the contributions of spin-torque, damping and thermal noise, a macrospin's dynamical evolution will deviate from a constant energy trajectory. Applied currents can reorient the magnetization by pumping energy into the magnetic system. We may then ask how the constant energy trajectories will be perturbed. This can be expressed mathematically by computing how the magnetization energy changes as a result of LLGS evolution. Taking the time derivative of~(\ref{eq:epsilon1}), we write~\footnote{The chain rule for stochastic variables is unchanged if the multiplicative noise follows the Stratonovich convention.}:

\begin{equation}
\label{eq:dotenergy}
\dot{\epsilon}=2\left[Dm_z\dot{m_z}-m_x\dot{m_x}\right]
\end{equation}
as the dynamical evolution equation for the macrospin's energy. Expressing the time derivatives of the magnetization components in terms of the full stochastic LLGS dynamics by using~(\ref{eq:Langevindynamics}), one obtains a stochastic evolution equation of the form

\begin{equation}
\label{eq:energy}
\dot{\epsilon}=f(\mathbf{m})+g(\mathbf{m})\circ\dot{W}.
\end{equation}

We now consider qualitatively how the macrospin dynamics change if the timescale for energy pumping/sinking, due to the collective effects of damping, spin-torque and thermal noise, is much larger than the precessional period of the conservative dynamics. In such a scenario, the full stochastic LLGS dynamics might be expected to follow constant energy trajectories fairly closely, with the macrospin drifting slowly from one constant energy trajectory to the other. Averaging the right hand side (RHS) of~(\ref{eq:energy}) over constant energy trajectories will then lead to a single stochastic differential equation for the evolution of the macrospin's energy. This approach is justified when the energy drift over the period of a single conservative orbit $T(\epsilon)\dot{\epsilon}$ be sufficiently small. As mentioned in the introduction,we focus here on deriving averaged energy dynamics valid in the domain $\epsilon>0$.~\footnote{Refer to our previous publication~\cite{PRB} for more details, and a discussion of the in-plane ($\epsilon<0$) precessional dynamics.}

In this approach, we now consider damping, applied current and thermal noise effects on an OOP $\epsilon>0$ orbit. First, we average (\ref{eq:energy}) over conservative positive energy trajectories (\ref{eq:posstates}). 
Due to the symmetry of such trajectories, most terms average to zero with the remaining nonzero terms leading to the constant-energy orbit-averaged (CEOA) equation:
\bea
\label{eq:eevolve}
\langle\partial_{\mathrm{t}}\epsilon\rangle&=&2\alpha\left[I(D-\epsilon)(\sin\omega\cos^2\psi)\langle m_z\rangle-D(D+1)\langle m_z^2\rangle+\epsilon(1+\epsilon)\right]+h(\epsilon)\nonumber\\
&+&\sqrt{\frac{2\alpha D(D+1)}{\xi}}\sqrt{\langle m_z^2\rangle-\frac{\epsilon(1+\epsilon)}{D(D+1)}}\cdot\dot{W_{\epsilon}},
\eea
where angular brackets $\langle\cdot\rangle$ denote averaging over a constant-energy trajectory with energy $\epsilon$. The second drift term (following the square brackets) $h(\epsilon)$ is a result of transforming~(\ref{eq:Langevindynamics}) into its It\={o} representation before performing the average over orbits (see Appendix B). As a result, the multiplicative noise terms appearing in the averaged energy equation above are now interpreted in the It\={o} sense\footnote{We distinguish equations written in It\={o} vs. Stratonovich form by writing the multiplicative noise as `$\cdot\dot{W_{\epsilon,\phi}}$'}.

We note that, as has been found for negative CEOA states~\cite{PRB}, the dynamics as a function of applied current for different spin-polarization tilts are identical, the current is simply rescaled by $\sin\omega\cos^2\psi$ (refer to Fig.~\ref{F1}). This allows us to numerically verify the CEOA approach by checking that the macrospin's evolution over some (properly rescaled) applied current is exactly identical for different tilts of the spin-polarization axes.

Under our assumptions, thermal noise will influence the dynamics in two distinct ways. The first, just discussed, is by nudging the magnetization onto a different energy orbit. The second, is by perturbing the precessional phase of the magnetization along a given constant energy orbit. As such,~(\ref{eq:eevolve}) must be supplemented by an equation describing the stochastic evolution of the dynamical phase. This can be written down by noting that noise must influence energy and phase diffusion identically because it is isotropic:

\be
\label{eq:phaseevolve}
\langle\partial_{\mathrm{t}}\chi\rangle=\frac{2\pi}{T(\epsilon)}+\sqrt{\frac{2\alpha D(D+1)}{\xi}}\sqrt{\langle m_z^2\rangle-\frac{\epsilon(1+\epsilon)}{D(D+1)}}\cdot\dot{W_{\chi}},
\ee 
where $T(\epsilon)$ is the period of the orbit at energy $\epsilon$. We distinguish between the two independent noise terms $\dot{W_{\epsilon}}$ and $\dot{W_{\chi}}$ by the fact that they act in orthogonal directions: respectively away and along the constant energy orbit. Whereas (\ref{eq:eevolve}) does not depend explicitly on the phase $\chi$, (\ref{eq:phaseevolve}) does however depend explicitly on the energy $\epsilon$. This will become important when we discuss different aspects of phase noise in Sec. VI.

To compute the averages $\langle m_z\rangle$ and $\langle m_z^2\rangle$ explicitly, we note that the positive energy trajectories can be geometrically parametrized as follows:

\bea
\label{eq:geoparam}
m_x^0(s)&=&\sqrt{\epsilon}\sinh(s),\\
m_y^0(s)&=&\pm\sqrt{1+\epsilon}\sqrt{1-\gamma^2\cosh^2(s)}\\
m_z^0(s)&=&\pm \sqrt{\frac{\epsilon}{D}}\cosh(s)\\
\gamma^2&=&\frac{\epsilon(D+1)}{D(\epsilon+1)},
\eea
where the parameter $s$ ranges from $-\acosh(1/\gamma)<s<\acosh(1/\gamma)$.
Upon computing the averages explicitly~(Appendix~A), the CEOA
equations for the positive energy dynamics 
($0<\epsilon<D$), expressed in terms of
$\gamma$, read~\footnote{We can allow ourselves the freedom to switch between
  expressions involving $\gamma$ and
  $\epsilon$. $\gamma^2=\frac{\epsilon(D+1)}{D(\epsilon+1)}$ is a
  monotonically increasing function of $\epsilon$ with the
  convenient property that $\epsilon=0\to\gamma=0$ and
  $\epsilon=D\to\gamma=1$. As such limits written in terms of
  $\gamma$ and $\epsilon$ are equivalent.}
\bea
\label{eq:edyn}
\partial_{\mathrm{t}}\epsilon(\gamma)&=&\frac{\pi\alpha}{\eta_0(\gamma)}\frac{D(D+1)}{[D(1-\gamma^2)+1]^{3/2}}\nonumber\\
&\times&\left\{\pm \tilde{I}(1-\gamma^2)-\frac{2}{\pi}\sqrt{D(1-\gamma^2)+1}\left[\eta_1(\gamma)-\frac{\gamma^2}{(D(1-\gamma^2)+1)}\eta_0(\gamma)\right]\right\}\nonumber\\
&+&h(\epsilon)\nonumber\\
&+&\sqrt{\frac{2\alpha}{\xi}\frac{D(D+1)}{D(1-\gamma^2)+1}\frac{1}{\eta_0(\gamma)}\left(\eta_1(\gamma)-\frac{\gamma^2}{D(1-\gamma^2)+1}\eta_0(\gamma)\right)}\cdot \dot{W_{\epsilon}}\\
\partial_{\mathrm{t}}\chi(\gamma)&=&\frac{\pi}{2\eta_0(\gamma)}\sqrt{\frac{D(D+1)}{D(1-\gamma^2)+1}}\nonumber\\
&+&\sqrt{\frac{2\alpha}{\xi}\frac{D(D+1)}{D(1-\gamma^2)+1}\frac{1}{\eta_0(\gamma)}\left(\eta_1(\gamma)-\frac{\gamma^2}{D(1-\gamma^2)+1}\eta_0(\gamma)\right)}\cdot \dot{W_{\chi}}
\eea
where $\eta_0(\gamma)=\mathrm{K}[1-\gamma^2]$ and $\eta_1(\gamma)=\mathrm{E}[1-\gamma^2]$ are expressed in terms of complete elliptic integrals of the first and second kind. For notational simplicity, the geometrical tilts have been absorbed into $\tilde{I}\equiv I\sin\omega\cos^2\psi$.\footnote{Note that in contrast to the in-plane precessional dynamics discussed in~\cite{PRB}, the current is rescaled by $\sin\omega$ as opposed to $\cos\omega$} It is important to note the applied current acts either to positively or negatively dampen the dynamics depending on which $\epsilon>0$ basin the magnetization is in (see Figure 2). The second drift term appearing on the third line of the RHS is the drift correction due to our change to It\={o} calculus. As discussed in Appendix B, the extra drift term results in a negligible correction. The following analysis will hence ignore its second order effects although they can be reintroduced straightforwardly if higher quantitative accuracy is desired.\footnote{The It\={o} drift-diffusion correction becomes relevant for dynamics close to the $\epsilon=0$ separatrix.}

In following the outlined procedure, we have reduced the
complexity of the magnetization dynamics to a
one-dimensional stochastic differential equation, whose properties we will now show to be analytically tractable. 

\section{Fixed Point Analysis}

As seen from~(\ref{eq:edyn}), in the absence of applied currents, the deterministic drift portion (first term on the RHS) of the energy diffusion dynamics is globally negative, $\partial_{\mathrm{t}}\epsilon<0$. The energy $\epsilon$ flows from positive to negative energy basins 
toward its minimum value of~$-1$. This is consistent with our physical notion of the $\epsilon>0$ basins being energetically unfavorable. Upon introducing an applied current, the behavior remains unchanged 
as long as no tilt is present between easy and spin-polarization axes ($\omega=0$). If a nonzero tilt is introduced into 
the system, the symmetry of the two positive energy basins is broken. In particular, due to the dependence on $\pm \tilde{I}$ (everything else inside the curly brackets is always negative), a critical current will exist, corresponding to a fixed point in the energy dynamics appearing in the positive $\hat{z}$, $\epsilon>0$ basin. The presence of a fixed point in the energy dynamics corresponds to a stable precessional (limit cycle) state of the magnetization dynamics. The dynamics in the negative $\hat{z}$ basin, on the other hand, will continue to be globally dissipative. Physically this is explained by the fact that the tilt $\omega$ biases the magnetic evolution away from one basin in favor of the other.

The critical current at which a fixed point appears can be obtained by studying the behavior of the energy dynamics in the limit $\epsilon=\gamma\to 0$. Requiring that

\be
\lim\limits_{\epsilon\to 0} \mathrm{T}(\epsilon)\dot{\epsilon}\propto -2\sqrt{D+1}+\pi \tilde{I}=0,
\ee
we obtain

\be
\tilde{I}_{\mathrm{OOP}}=\frac{2}{\pi}\sqrt{D+1},
\ee
as the current where a stable fixed point appears at $\epsilon=0$. Increasing $\tilde{I}$ further will shift the fixed point to higher energies.
Qualitatively, this will result in an increase of frequency and decrease of amplitude of the limit cycle oscillations. The maximum possible energy obtainable by 
the oscillator is $\epsilon=D$. This is achieved when~\footnote{One analogously seeks a null net drift of the energy dynamics at $\epsilon=D$: $\lim\limits_{\epsilon\to D}\mathrm{T}(\epsilon)\dot{\epsilon}=0$}

\be
\tilde{I}_{\mathrm{max}}=D+\frac{1}{2}.
\ee
Increasing the current beyond $\tilde{I}_{\mathrm{max}}$ simply overdrives the magnetization. As we will see later, the CEOA approximation breaks down beyond this point and stable oscillations disappear. Fig.~4 shows a sample of the drift field due to~(\ref{eq:edyn}) for $\tilde{I}<\tilde{I}_{\mathrm{OOP}}$, $\tilde{I}_{\mathrm{OOP}}<\tilde{I}<\tilde{I}_{\mathrm{max}}$, and $\tilde{I}>\tilde{I}_{\mathrm{max}}$. $\tilde{I}_{\mathrm{OOP}}$ and $\tilde{I}_{\mathrm{max}}$ represent the lower and upper threshold currents for the appearance of steady-state precessions in the stable OOP basin due to the nonlinear character of the magnetization dynamics.

Comparing with the CEOA treatment of magnetic switching~\cite{PRB}, we note that $\tilde{I}_{\mathrm{switch}}$, the critical current for switching, equals $\sqrt{D}\tilde{I}_{\mathrm{OOP}}$. As such, the minimal currents sustaining stable OOP precessional states are generally smaller than the critical switching current. This results in the prediction of a hysteretic dependence of IP$\rightleftharpoons$OOP transitions on applied current, which has been observed recently in experiment~\cite{Li}. In detail, since $\tilde{I}_{\mathrm{switch}}=I_{\mathrm{switch}}\cos\omega$ and $\tilde{I}_{\mathrm{OOP}}=I_{\mathrm{OOP}}\sin\omega\cos^2\psi$, one can see that the relation between direct critical switching current and threshold current for sustainment of OOP precessions is

\be
\label{eq:critcur}
I_{\mathrm{OOP}}=\frac{I_{\mathrm{switch}}}{\sqrt{D}\tan\omega\cos^2\psi}.
\ee

\begin{figure}
	\begin{center}
	\centerline{\includegraphics [clip,width=5in, angle=0]{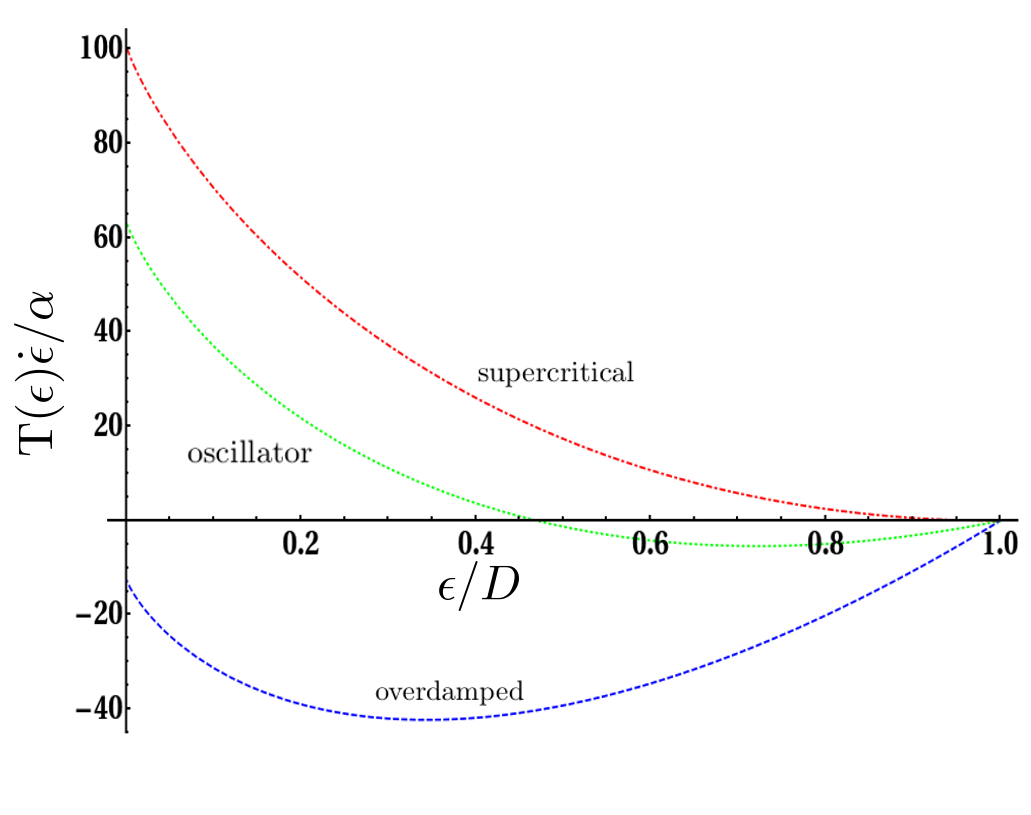}}
	\caption{Three regimes of deterministic energy flow $\dot{\epsilon}$ as a function of energy for $D=10$. (blue-dashed) $\tilde{I}<\tilde{I}_{\mathrm{OOP}}$: Subcritical regime. Energy flows from positive to negative energy basins due to dynamics being globally dissipative (overdamped). (red-dashdotted) $\tilde{I}>\tilde{I}_{\mathrm{max}}$: Supercritical regime. Energy flows towards limiting stable value $\epsilon=D$ due to dynamics being overdriven by applied current. (green-dotted) $\tilde{I}_{\mathrm{OOP}}<\tilde{I}<\tilde{I}_{\mathrm{max}}$: Oscillator regime. Energy flow will stabilize at a fixed point corresponding to a precessing oscillator state. In this regime, the fixed point represents a constant energy trajectory where spin-torque and damping effects balance.}
	\label{F4}
	\end{center}
\end{figure}

\section{Limits of the CEOA Approach}

For our approximations to be valid, the averaged energy flow
($T(\epsilon)|\partial_t\epsilon|$) over any given orbit must be small
compared to the maximum allowable energy variations ($0<\epsilon<D$):

\be
\label{eq:validity}
\underaccent{\epsilon}{\mathrm{max}}\: T(\epsilon)\lvert\partial_{\mathrm{t}}\epsilon\rvert\ll D.
\ee
This has been discussed elsewhere~\cite{PRB,TaniguchiE} so we simply state the results for OOP dynamics. For CEOA to be applicable one must have $\tilde{I}_{\mathrm{OOP}} \lesssim \tilde{I} \lesssim \tilde{I}_{\mathrm{max}}$. 

In Fig.~5 we show a comparison between theory and numerical results by plotting average energy $\langle\epsilon\rangle$ as a function of applied current. Ensembles consisting of 10000 macrospins were initialized antiparallel to the easy-axis and allowed to relax subject to a steady applied current.  Upon varying the angular tilt $\omega$ between easy and spin-polarization axes, we notice that the data follow our theory down to a minimum critical angle $\omega_C$. For angular tilts less than $\omega_C$, stable positive energy steady states cease to be accessible regardless of the applied current. The origin of this angular cutoff is geometrical in nature and corresponds to the necessity for the spin-polarization axis to be pointing inside the positive energy basin. The condition for this to happen can be seen from~(\ref{eq:epsilon1}) by solving for the separatrix of the energy basins. One obtains

\be
\label{eq:crittilt}
\omega_C=\frac{\pi}{2}-\mathrm{arctan}(\sqrt{D}),
\ee     
which is in excellent agreement with numerical data. This geometrical intuition can be seen from theory by determining the tilt for which the threshold current for OOP precessions equals that for direct switching. Starting with~(\ref{eq:critcur}), and setting $\psi=0$ for convenience, leads to~(\ref{eq:crittilt}).

\begin{figure}
	\begin{center}
	\centerline{\includegraphics [clip,width=7in, angle=0]{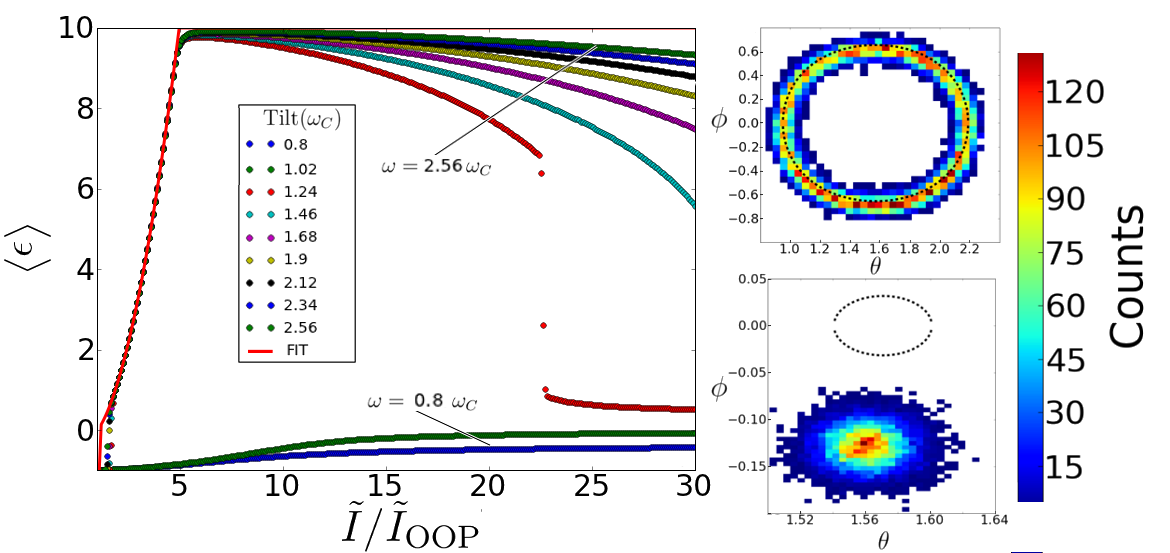}}
	\caption{Steady-state ensemble energy as a function of dimensionless applied current $\tilde{I}$ (rescaled by $\tilde{I}_{\mathrm{OOP}}=(2/\pi)\sqrt{D+1}$) for a model with $D=10$, $\xi=80$ and $\alpha=0.04$. Red line shows an analytic fit to numerical data within the current limits defined by the theory (for reference $\tilde{I}_{\mathrm{max}}/\tilde{I}_{\mathrm{OOP}}\approx 4.97$). Insets shows density plots in spherical coordinates of 10000 numerical trajectories for a sample with a $ 2.56 \omega_C$ tilt between easy and spin-polarization axes, driven by a current of $\tilde{I}/\tilde{I}_{\mathrm{OOP}}=4$ (top), and $\tilde{I}/\tilde{I}_{\mathrm{OOP}}=15$ (bottom). The dotted line denotes the conservative trajectory.}
	\label{F5}
	\end{center}
\end{figure}

For large currents $\tilde{I}>\tilde{I}_{\mathrm{max}}$, numerical results seem to indicate a steady drop in ensemble energy as the applied current is increased. In fact, contrary to the CEOA description, the macrospin's magnetization ceases to precess around the hard axis and instead settles into a magnetic configuration where all static torques balance and spin-torque effects compete with the magnetic anisotropies.

\section{Thermal Stability, Precession Linewidth, Phase, Amplitude and Power Fluctuations}

So far, we have provided an analytical approach that enables the study of the properties of OOP dynamics. Once the strength of the applied current $\tilde{I}$ has been chosen, and provided that the angular tilt of the spin-polarization vector is sufficient ($\omega>\omega_C$), the average energy $\epsilon_0=\langle\epsilon\rangle$ of the equilibrium steady state trajectory can be obtained by solving for the fixed point of the energy dynamics~(\ref{eq:edyn}). Due to the dependence of the precessional period $\mathrm{T}(\epsilon)$ on the energy of the orbit, the expected precessional frequency can be inferred. 

Thermal noise will, however, perturb the magnetization about the fixed point, resulting in fluctuations of the macrospin's energy around its average $\epsilon_0$ value and diffusion of its phase $\chi$ along the relevant constant energy orbit. These deviations are believed to be the source of the oscillator's experimentally measured frequency, linewidth and phase decoherence. We will now proceed to derive an estimate for such linewidths.

The general stochastic energy evolution equation~(\ref{eq:edyn}) can be written concisely as

\be
\label{eq:dynred}
\partial_t\epsilon=f(\epsilon,\tilde{I})+h(\epsilon)+g(\epsilon)\cdot\dot{W_{\epsilon}},
\ee 
where $f(\epsilon,\tilde{I})$, $h(\epsilon)$ and $g(\epsilon)$ are, respectively, the deterministic drift, It\={o} drift-diffusion correction and multiplicative noise. Following Ref.~[22], one can use the stochastic energy evolution equation to compute the mean time one must wait to observe a thermal excitation out of an OOP trajectory. The asymptotic dependence of such a mean escape time is then

\be
\log(\langle\tau_\mathrm{jump}\rangle)\propto 2\int_0^{\epsilon_0(\tilde{I})}dx\frac{f{(x,\tilde{I})}}{g^2(x)}=\xi\left(\epsilon_0-\frac{\tilde{I}}{\tilde{I}_{\mathrm{OOP}}}\int_0^{\epsilon_0}dx\frac{D-x}{\sqrt{1+x}(D\eta_1(x)-x\eta_0(x))}\right),
\ee
where $\epsilon_0\equiv\langle\epsilon\rangle$ is the usual solution of the fixed point equation (dependent on $\tilde{I}$). Due to the dependence of the equilibrium oscillator energy on the applied current $\epsilon_0(\tilde{I})$, the thermal stability of the OOP precessional states will depend non-linearly on the applied current $\tilde{I}$.

The Fokker-Planck (FP) equation is:
\be
\partial_t\rho=\partial_{\epsilon}\left[f(\epsilon,\tilde{I})\rho-\frac{1}{2}g^2(\epsilon)\partial_{\epsilon}\rho\right],
\ee
whose solution describes the full evolution of the energy distribution $\rho(\epsilon,t)$ as a function of time (Appendix B). At equilibrium ($\partial_t\rho=0$), the saddle point approximation can be used to determine a steady state distribution
\be
\label{eq:equidist}
\rho_{\mathrm{eq}}(\epsilon)\propto \exp\left[2\int_0^\epsilon dx\frac{f(x,\tilde{I})}{g^2(x)}\right]\simeq\exp\left[\frac{f'(\epsilon_0,\tilde{I})}{g^2(\epsilon_0)}(\epsilon-\epsilon_0)^2\right],
\ee
that is valid as long as $\tilde{I}>\tilde{I}_{\mathrm{OOP}}$. We can then write an expression for the amplitude noise by computing the variance of the energy in an equilibrium OOP distribution:

\be
\label{eq:Evar}
\langle(\epsilon-\epsilon_0)^2\rangle\simeq\frac{g^2(\epsilon_0)}{2\lvert f'(\epsilon_0,\tilde{I})\rvert}.
\ee

In Fig.~6 we compare the theoretical approximation resulting from~(\ref{eq:Evar}) with the equilibrium energy variance extracted from our numerical simulations. Whereas the variance does not appear to rescale trivially with the spin-polarizer tilt, all tilts seem to show a variance versus applied current curve that peaks within the same general region predicted by our rough estimate. For currents $\tilde{I}\simeq \tilde{I}_{\mathrm{OOP}},\tilde{I}_{\mathrm{max}}$ the approximation breaks down due to failure of the CEOA approximation.  

\begin{figure}
	\begin{center}
	\centerline{\includegraphics [clip,width=4in, angle=0]{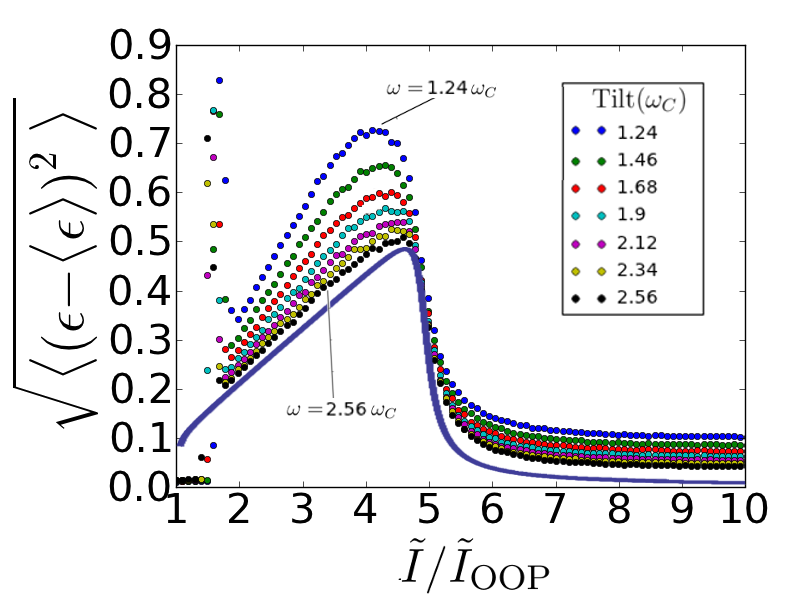}}
	\caption{Standard deviation of the energy distribution plotted as a function of dimensionless applied current $\tilde{I}$ (rescaled by $\tilde{I}_{\mathrm{OOP}}=(2/\pi)\sqrt{D+1}$) for $D=10$, $\xi=80$ and $\alpha=0.04$. The solid blue line shows the theoretical prediction~(\ref{eq:Evar}) calculated within the current limits defined by the theory (for reference $\tilde{I}_{\mathrm{max}}/\tilde{I}_{\mathrm{OOP}}\approx 4.97$).}
	\label{F6}
	\end{center}
\end{figure}

Using~(\ref{eq:equidist}), we see that all energy-dependent stationary characteristics $\langle Q\rangle=\int Q(\epsilon)\rho_{\mathrm{eq}}(\epsilon)$ of the oscillator can be computed via distribution averaging. However, we employ our saddle point estimate~(\ref{eq:Evar}) to study thermal fluctuations. The relative fluctuation of a quantity $Q(\epsilon)$ at equilibrium will be given by $\delta Q/Q=(Q'(\epsilon)/Q(\epsilon))\vert_{\epsilon=\epsilon_0}\sqrt{\langle(\epsilon-\epsilon_0)^2\rangle}$. 

As a first example, the experimentally observed oscillator power depends on the square of the oscillator's precession amplitude along the in-plane direction. Having chosen a coordinate system with the reference magnetic layer aligned in-plane, power fluctuations are directly proportional to fluctuations in the precession amplitude of the oscillator as projected along the in-plane axial direction. From our previously derived expression of the oscillation amplitude along the in-plane direction~(\ref{eq:Zamplitude}), one has:

\be
\frac{\delta \mathrm{P}}{\mathrm{P}}=\frac{\delta A_{\mathbf{\hat{x}}}^2}{A_{\mathbf{\hat{x}}}^2}\simeq\sqrt{\langle(\epsilon-\epsilon_0)^2\rangle}.
\ee 

\begin{figure}
	\begin{center}
	\centerline{\includegraphics [clip,width=4in, angle=0]{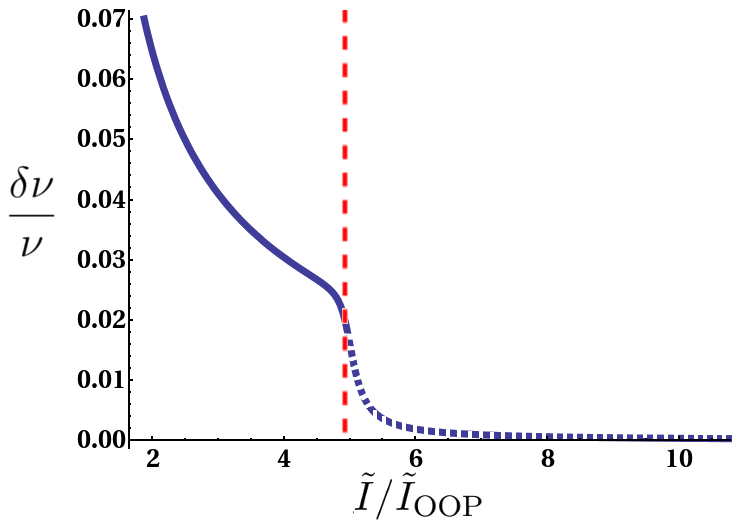}}
	\caption{Inverse quality factor~(\ref{eq:qual}) vs. applied current for $D=10$ set at room temperature ($\xi=80$). Red dashed line denotes the upper bound of the validly of the CEOA formalism: $\tilde{I}_{\mathrm{max}}/\tilde{I}_{\mathrm{OOP}}\approx 4.97$ for the parameters chosen.}
	\label{F7}
	\end{center}
\end{figure}

Analogously, denoting the oscillation frequency by $\nu(\epsilon)=2\pi/T(\epsilon)$, one finds for the precession linewidth quality factor $Q$ dependence on amplitude noise:

\be
\label{eq:qual}
\frac{1}{Q}=\frac{\delta\nu}{\nu}\simeq\frac{\mathrm{T}'(\epsilon_0)}{\mathrm{T}(\epsilon_0)}\sqrt{\langle(\epsilon-\epsilon_0)^2\rangle}.
\ee  
Fig.~\ref{F7} shows how the quality factor is a monotonically increasing function of applied current. Overall, increasing the driving current reduces the linewidth of the oscillator in line with classical oscillator theory which predicts a linewidth scaling dependent on the ratio of the thermal and oscillator energy ($k_BT/\epsilon$). In practice, however, at currents high enough for the breakdown of the macrospin model, micromagnetic effects due to Oersted fields are expected to complicate the physical picture in non-trivial ways.

One may proceed further and ask whether the CEOA formalism is capable of shedding light on the phase noise and, more generally, the phase decoherence driving a magnetic system. The assumption that ``sufficiently weak" noise drives diffusion from one energy orbit to another does not impose any limit on how strong the noise driving the phase of the actual constant energy oscillation can be. Both phase noise due to thermal diffusion along a given constant energy orbit and amplitude noise can drive phase decoherence in a magnetic system. As such, the relative intensity of both effects must be determined to understand phase decoherence.
 

To do so, we consider how energy fluctuations about the $\epsilon_0$ equilibrium fixed point influence the phase dynamics described in (\ref{eq:edyn}). Let $\epsilon(t)\equiv\epsilon_0+\delta\epsilon(t)$ and expand (\ref{eq:dynred}) in powers of $\delta\epsilon$. Denoting $F(\epsilon)\equiv f(\epsilon)+h(\epsilon)$, the resultant stochastic differential equation can be formally integrated to give:
 
\be
\label{eq:efluc}
\delta\epsilon(t)=e^{F'(\epsilon_0)t}\left[c+g(\epsilon_0)\int_0^tdt'e^{-F'(\epsilon_0)t'}\cdot \dot{W}_{\epsilon}\right],
\ee
where primes represent differentiation with respect to energy ($F'(\epsilon_0)\equiv\partial_{\epsilon}F|_{\epsilon=\epsilon_0}$), and $c$ is an (unimportant) initial condition. $|F'|$ represents the relaxation rate of amplitude fluctuations to the $\epsilon_0$ baseline. Given the explicit dependence of the phase $\chi$ on the energy evolution, such energy fluctuations are expected to play a crucial role in the thermally driven phase dynamics. 

Expanding the phase dynamics about $\epsilon_0$ to lowest order, we have:

\be
\label{eq:eflucphase}
\partial_{\mathrm{t}}\chi=\frac{2\pi}{T(\epsilon_0)}-\frac{2\pi T'(\epsilon_0)}{T^2(\epsilon_0)}\delta\epsilon(t)+g(\epsilon_0)\cdot \dot{W}_{\chi}.
\ee
Substituting (\ref{eq:efluc}) into (\ref{eq:eflucphase}) and recalling that $\dot{W}_{\epsilon}$ and $\dot{W}_{\chi}$ are uncorrelated stochastic processes, the expected phase variance at equilibrium can be evaluated to give (we suppress the dependence on $\epsilon_0$):

\be
\label{eq:phasevar}
\langle\Delta\chi^2\rangle(t)=g^2\left\{\left[1+\left(\frac{2\pi T'}{F'T^2}\right)^2\right]|t|+\frac{1}{2F'}\left(\frac{2\pi T'}{F'T^2}\right)^2\left[4\left(1-e^{F'|t|}\right)-\left(1-e^{2F'|t|}\right)\right]\right\}.
\ee
which closely resembles the more general prediction from oscillator theory.\cite{STIEEE08,JVKBook} Since the power spectrum can be written as a Fourier transformation of the correlation function $\langle\exp[i(\chi(t)-\chi(t))]\rangle\approx\exp[i\langle\chi(t)-\chi(t)\rangle]\exp[-\langle\Delta\chi^2\rangle(t)/2]$, the linewidth can be predicted\cite{TaniSolo} by inspecting (\ref{eq:phasevar}).

The temporal dependence of the phase variance is responsible for the decoherence of the magnetic ensemble over time. We interpret the {\it decoherence time} $\tau_{\mathrm{dec}}$ as the timescale necessary for the ensemble to homogeneously distribute itself along a given constant energy orbit similarly to what is shown in Fig. 5. We quantify $\tau_{\mathrm{dec}}$ by asking on what timescale the width of the phase distribution begins to encompass the entire constant energy orbit: $\langle\Delta\chi^2\rangle(\tau_{\mathrm{dec}})=4\pi^2$.  Although the temporal dependence is generally quite complicated, two limiting regimes can be explored. For low enough temperatures, the phase decoherence time $\tau_{\mathrm{dec}}$ will be larger than the relaxation timescale of the amplitude fluctuations $\tau_{\mathrm{dec}}\gg 1/|F'|$. Decoherence can then be expected to mostly take place due to the differences in orbital evolution at the different energies explored by the amplitude fluctuations. This will eventually lead the spin ensemble to decohere and thermalize to a homogenous distribution of phases relative to the referential $\epsilon_0$ orbit. The dominant amplitude fluctuations driving such a low temperature regime result in a linear dependence of the phase variance.

\be
\langle\Delta\chi^2(t)\rangle\approx g^2\left[1+\left(\frac{2\pi T'}{F'T^2}\right)^2\right]|t|.
\ee
Due to the dependence of the multiplicative noise term in~(\ref{eq:edyn}) on temperature ($g(\epsilon)\propto\sqrt{T}$), the decoherence time $\tau_{\mathrm{dec}}\propto T^{-1/2}\propto\sqrt{\xi}$ can be predicted to depend on the inverse square root of temperature. Furthermore, a linear dependence on time will imply a Lorentzian power spectrum with linewidth $\Delta \nu_L=(g^2/2\pi)(1+\mu^2)$ ($\mu=2\pi T'/F'T^2$).

In a high temperature limit, pure phase noise will compete with the amplitude noise effects by decohering the ensemble on a timescale smaller than the amplitude fluctuation relaxation rate $\tau_{\mathrm{dec}}\ll 1/|F'|$. The exponential contributions in (\ref{eq:phasevar}) cease to be negligible and the approximate temporal dependence of the phase variance can be written to second order in time as:

\be
\langle\Delta\chi^2(t)\rangle\approx g^2\left[|t|+2\left(\frac{2\pi T'}{F'T^2}\right)^2|F'||t|^2\right].
\ee
If $(2\pi T'/\sqrt{|F'|}T^2)^2\gg 1$ (typically the case when $\epsilon_0\ll D$), the term linear in time can be dropped altogether resulting in a purely quadratic dependence of the phase variance on time. In such a scenario, the decoherence time can be expected to scale linearly with the inverse temperature $\tau_{\mathrm{dec}}\propto T^{-1}\propto\xi$. A phase variance scaling quadratically in time will in turn lead to a gaussian power spectrum with linewidth $\Delta \nu_L=\sqrt{2g\mu^2F'}/2\pi$.

\begin{figure}
	\begin{center}
	\centerline{\includegraphics [clip,width=4.3in, angle=0]{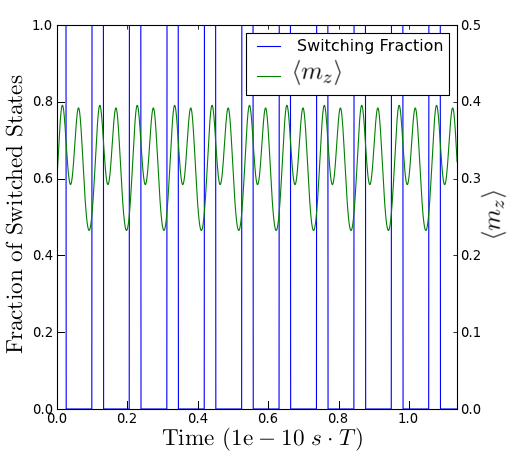}}
	\caption{Switching probability vs. spin-current pulse length for a macrospin model with $D=10$, $\omega=2.12\,\omega_C$) driven by a spin-current intensity of $\tilde{I}=2.75\,\tilde{I}_{\mathrm{OOP}}$ in the absence of thermal noise. Times are shown in units of ($s\cdot T$) where $T$ stands for Tesla: real time is obtained upon division by $\mu_0 H_K$. Before the current pulse is switched on, the magnetic ensemble is taken to be antiparallel to the easy-axis of the magnetic film. Switching probability is defined as the ensemble fraction that relaxes into a parallel configuration upon switching the current pulse off. The right-hand vertical axis plots the evolution of the average $\langle m_z\rangle$ component. In the absence of thermal noise the oscillator remains coherent at all times and its periodic motion is clearly seen. Due to the deterministic nature of the zero-temperature dynamics, the macrospin will deterministically switch either into the parallel or antiparallel state at all times.}
	\label{F8}
	\end{center}
\end{figure}
\begin{figure}
\centering
\begin{minipage}[t]{.45\textwidth}
\centering
\includegraphics [clip,width=3.7in, angle=0]{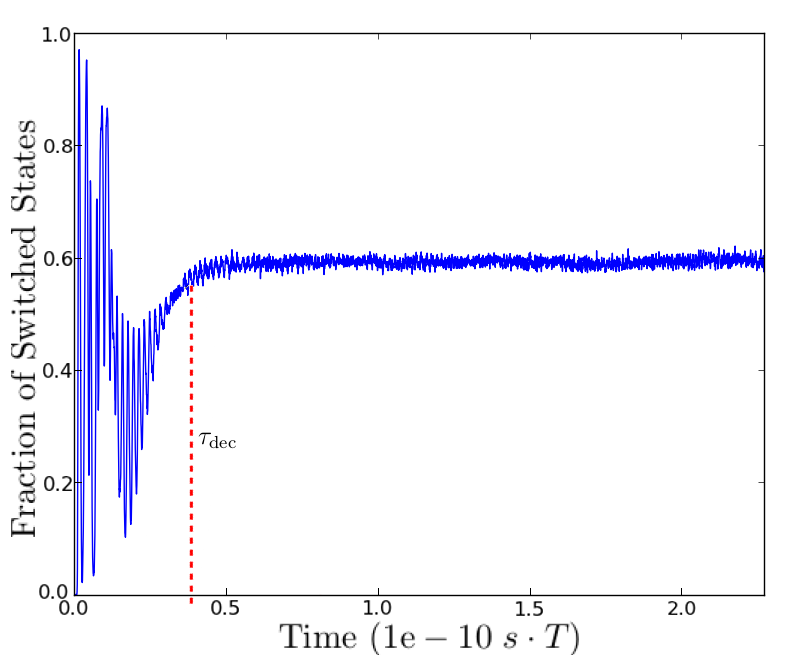}
\end{minipage}\hfill
\begin{minipage}[t]{.45\textwidth}
\centering
\includegraphics [clip,width=3.4in, height=3.04in, angle=0]{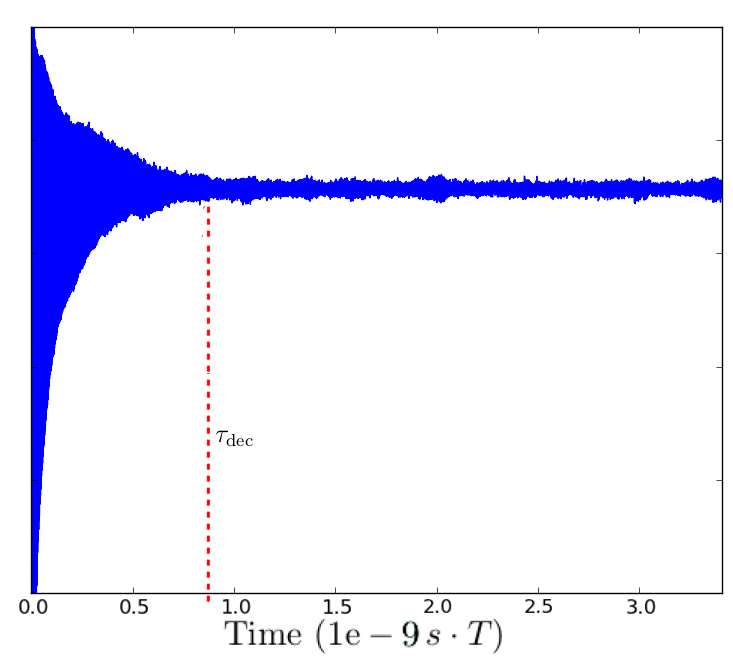}
\end{minipage}
\caption{Switching probability vs. spin-current pulse length for a macrospin model with $D=30$, $\omega=3\omega_C$ driven by a spin-current intensity of $\tilde{I}=5 \,\tilde{I}_{\mathrm{OOP}}$ in the presence of thermal noise corresponding to $\xi=80$ (left) and $\xi=1200$ (right). Times are shown in units of ($s\cdot T$) where $T$ stands for Tesla: real time is obtained upon division by $\mu_0 H_K$. Before the current pulse is switched on, the magnetic ensemble is taken to be antiparallel to the easy-axis of the magnetic film. Switching probability is defined as the fraction of the ensemble that relaxes into a parallel configuration upon switching the current pulse off. For long pulse times the switching probability converges to a value indicating that the phase of the OOP precession has decohered. The red dashed lines are a qualitative graphical representation of the decoherence time.}
\label{F9}
\end{figure}

We explore these predictions by studying switching probability curves of a macrospin ensemble at varying temperatures for applied current intensities and effective spin-polarization axial tilt consistent with an OOP precessional behavior. Upon switching the current off, the phase of the oscillator will select the macrospin's relaxation outcome (either parallel or antiparallel to the easy axis of the magnetic film) with high probability. In the absence of thermal noise, a current pulse of fixed duration will lead to either a parallel or antiparallel relaxed state after the pulse terminates (see Fig.~\ref{F8}) with absolute certainty. At nonzero temperatures, however, oscillator ensemble phase decoherence is expected due to thermal noise. As a result, long spin-current pulse times will lead to equally likely parallel (antiparallel) relaxation due to ensemble thermalization along the OOP constant energy orbit. In Fig.~\ref{F9} we find good qualitative agreement between such an understanding of phase decoherence behavior and numerical simulations. The equilbrium probability bias for higher P switching is due to some of the states thermally equilibrating into the IP energy basin before the current pulse is switched off. 

\begin{figure}
	\begin{center}
	\centerline{\includegraphics [clip,width=4.3in, angle=0]{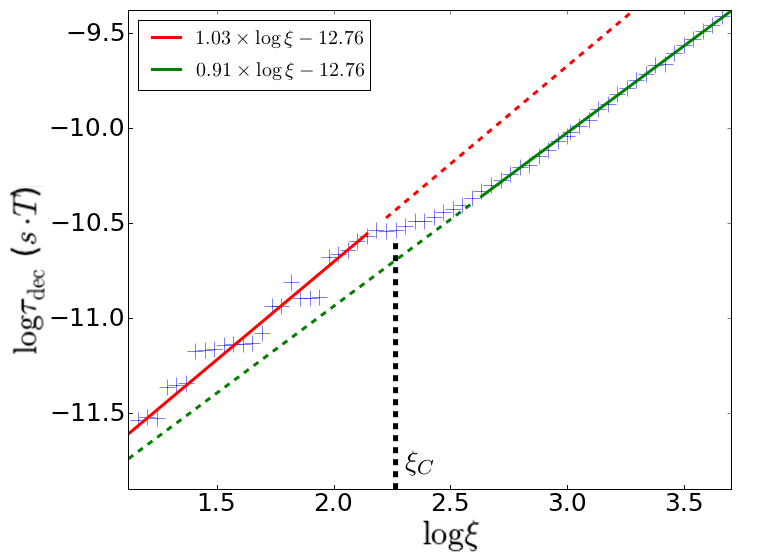}}
	\caption{Log-log plot of ensemble decoherence time vs. energy barrier height to thermal energy ratio $\xi$ for a macrospin model with $D=30$, $\omega=3\,\omega_C$ driven by a spin-current intensity of $\tilde{I}=1.5\,\tilde{I}_{\mathrm{switch}}$. Times are shown in units of ($s\cdot T$) where $T$ stands for Tesla: real time is obtained upon division by $\mu_0 H_K$. Linear regression (solid lines) of data points demonstrates a transition between a phase noise dominated regime $\tau_\mathrm{dec}\propto1/T$ below a certain critical inverse temperature $\xi<\xi_C$. Above $\xi>\xi_C$ ($T<T_C$), both amplitude and phase noise contribute to ensemble decoherence.}
	\label{F10}
	\end{center}
\end{figure}

The switching probability curves can be employed to numerically extract the decoherence time at different temperatures. Fig.~\ref{F10} shows a log-log plot of $\tau_\mathrm{dec}$ on $\xi$ for a $D=30$ model with a $\omega=3\,\omega_C$ tilt, driven by a $\tilde{I}=1.5\,\tilde{I}_{\mathrm{switch}}$ applied current. Linear regression to numerical data shows an inverse proportionality $\tau_\mathrm{dec}\propto 1/T\propto\xi$ between decoherence time and temperature for temperatures larger than a certain critical temperature. For $T<T_C$, however, both amplitude and phase noise seem to contribute to ensemble decoherence thus not allowing us to probe the pure amplitude noise decoherence mechanism previously discussed. 

\section{Conclusion}

We have analyzed the out-of-plane (OOP) precessional behavior of a biaxial macrospin in the presence of spin-torque due to both a perpendicularly magnetized polarizer and an in-plane (IP) magnetized reference layer. Their combined spin-torque effects lead to an effective tilt $\omega$ between the easy- and spin-polarization axes. The problem was treated analytically by employing multiscaling techniques to separate the fast oscillatory behavior due to conservative dynamical terms from the slow magnetic diffusion due to noise and spin-transfer torque. By averaging the stochastic LLG dynamics over constant energy trajectories we constructed a 1D stochastic evolution equation for the macrospin's energy. 
The features of the energy evolution equation were explored in detail analytically, and confirmed by numerically simulating the full thermally activated LLGS dynamics. We found that our multiscaling assumptions are valid for normalized applied currents in the range $(2/\pi)\sqrt{D+1}<I\sin\omega\cos^2\psi<D+1/2$, where $D$ is the ratio between hard- and easy-axis anisotropy, $I$ a rescaled applied current and $\omega$ the effective tilt between easy- and spin-polarization axes.

Within this regime, we found that changing the effective tilt serves to rescale the applied current; the dynamical behavior is otherwise identical. For applied currents greater than $I_{\mathrm{OOP}}=(2/\pi)\sqrt{D+1}/\sin\omega$, a stable fixed point appears in the macrospin's energy dynamics. This is consistent with the description of a stable limit cycle, interpreted as an OOP precessional state. We predict that stable OOP precessions are possible only in one of the two out-of-plane directions, selected by the direction of the applied current. Furthermore, by comparing our results to those obtained via CEOA methods to study the threshold currents for magnetic switching, we predict the occurence of hysteretic transitions between IP and OOP stable states for effective tilts larger than a critical tilt $\omega_C=\mathrm{arctan}(1/\sqrt{D})$, which has been observed in very recent experiments\cite{Li}. For tilts $\omega<\omega_C$, we predict that magnetic switching will take place since the threshold current for onset of stable OOP precessionary states is expected to be larger than that required for a direct switch. Overall, this leads to a very simple condition that a spin-valve must satisfy to behave like a STNO ($\eta_{\mathrm{ref}}/\eta_{\mathrm{pol}}<\sqrt{D}$). Our theory agrees with numerical results and could be a starting point for testing how well the macrospin approximation captures the magnetization dynamics in real devices. 

Upon exploring the thermal contribution to oscillator linewidth broadening, we observe the existence of a critical temperature $T_C$ separating a regime where phase noise dominates decoherence and one where decoherence is the result of both phase and amplitude noise. The former cannot be accounted for by our CEOA theory and is a result of the full complexity of the LLG dynamics. This is in agreement with the non-linear oscillator model where a transition temperature is predicted to exist between a phase noise dominated regime at large temperatures and one limited by thermal deflections about the equilibrium magnetic trajectory at low temperatures~\cite{Sankey, JVKBook}.

Our methodology is similar to that proposed by Slavin, Tiberkevich and Kim~\cite{JooVonPRL,STIEEE,STIEEE08}. However, instead of approaching the multiscaling analysis by studying the complex oscillatory amplitude of the macrospin's dynamics using a self-oscillator equation, we focused on the macrospin's diffusion over its energy landscape. The loss of generality in doing so is compensated by new insights into the macrospin's dynamical characteristics.

\section*{Acknowledgments}

  The authors would like to acknowledge J.-V. Kim, E. Vanden-Eijnden, K. Newhall, A. MacFadyen and J. Z. Sun for
  useful discussions and comments leading to this paper. This research
  was supported by NSF-PHY-0965015, NSF-DMR-100657 and NSF-DMR-1309202.

\appendix
\section{$\langle m_z\rangle$ and $\langle m_z^2\rangle$}

To compute the constant energy orbit averages in~(20) we write the integrals using the geometric parametrization~(21-24):

\bea
\langle m_z\rangle_{\bf{m}^0}=\frac{\pm}{\mathrm{T}(\epsilon)}\int_0^\mathrm{T}dtm_z(t)&=&\frac{\pm4}{\mathrm{T}(\epsilon)}\int_0^{\mathrm{acosh(1/\gamma)}}ds\lvert\frac{\partial_sm_z^0}{\dot{m}_z^0}\rvert m_z^0\nonumber\\
&=&\frac{\pm 4}{\mathrm{T}(\gamma)}\frac{\gamma}{\sqrt{D(D+1)}}\int_0^{\mathrm{acosh(1/\gamma)}}ds\frac{\cosh(s)}{\sqrt{1-\gamma^2\cosh^2(s)}}\nonumber\\
&=&\frac{\pm \pi}{2\sqrt{D(1-\gamma^2)+1}}\frac{1}{\mathrm{K}[1-\gamma^2]}.
\eea

Proceeding analogously for $\langle m_z^2\rangle$:

\bea
\langle m_z^2\rangle_{\bf{m}^0}=\frac{1}{\mathrm{T}(\epsilon)}\int_0^\mathrm{T}dtm_z^2(t)&=&\frac{4}{\mathrm{T}(\epsilon)}\int_0^{\mathrm{acosh(1/\gamma)}}ds\lvert\frac{\partial_sm_z^0}{\dot{m}_z^0}\rvert(m_z^0)^2\nonumber\\
&=&\frac{4}{\mathrm{T}(\gamma)}\frac{\gamma^2}{\sqrt{D(D+1)}\sqrt{1+D(1-\gamma^2)}}\int_0^{\mathrm{acosh(1/\gamma)}}ds\frac{\cosh^2(s)}{\sqrt{1-\gamma^2\cosh^2(s)}}\nonumber\\
&=&\frac{1}{1+D(1-\gamma^2)}\frac{\mathrm{E}[1-\gamma^2]}{\mathrm{K}[1-\gamma^2]},
\eea
where, as stated in the main text, $\mathrm{E}[x]$ is the complete elliptic integral of the second kind.

In both derivations we have taken advantage of eqns. (15) and (24) to write the period as a function of $\gamma$. Written explicitly, the period reads:

\be
\mathrm{T}(\epsilon)=\frac{4}{\sqrt{D(1+\epsilon)}}\mathrm{K}[\frac{D-\epsilon}{D(1+\epsilon)}]=4\sqrt{\frac{1+D(1-\gamma^2)}{D(D+1)}}\mathrm{K}[1-\gamma^2].
\ee

\section{Orbit averaging of a Stratonovich Equation}

There are several advantages in adopting a Stratonovich convention when writing the dynamical equations. First, it is the most natural way of modeling a physical process where the Gaussian noise represents the short correlation time limit of a colored noise process: by the Wong-Zakai theorem\cite{WongZakai}, such a limit of multiplicative noise converges to Statonovich calculus. Second, a Stratonovich interpretation follows the conventional rules of calculus in dealing with functions of a stochastic variable. Third, many conventional numerical schemes used to simulate Langevin equations (such as the Heun scheme adopted for this work) evolve towards the Stratonovich solution.

The Stratonovich formulation of a stochastic differential equation (SDE), however, fails to accurately represent the correlation between multiplicative terms and the specific noise realization~\cite{Karatsas}. To average the multiplicative noise terms over constant energy orbits, we take advantage of the fact that sums of Gaussian random variables $\sum_i \mu_ix_i$ (where $x_i$ are standard $0$ mean and variance $1$ Gaussian variables) behave like a single Gaussian variable $\tilde{x}$ with variance given by the square sum of the individual variances $\tilde{\mu}^2=\sum_i\mu_i^2$. Since the multiplicative noise terms $\hat{\mathbf{B}}(\mathbf{m})\circ\dot{\mathbf{W}}$ appearing in our LLGS equations are state-dependent, the Gaussian variable summation cannot be employed due to the temporal correlation between the state-dependent variances $\hat{\mathbf{B}}^2(\mathbf{m})$ and the specific noise realization $\dot{\mathbf{W}}$.

This problem can be avoided by converting the LLGS equations into their It\={o} representation. The multplicative noise terms of~(\ref{eq:eevolve}) become $(Dm_z\hat{B}_{xj}-m_x\hat{B}_{zj})\cdot\dot{W}_j$ (with summation over repeated indices). The state-dependent variances are now uncorrelated with respect to the noise realization, and so a summation of Gaussian random variables can now be employed. Averaging over constant energy orbits then leads, after a bit of algebra, to the noise term appearing in~(\ref{eq:eevolve}). 

Altering the multiplicative noise convention can generally alter the qualitative nature of the solution to the stochastic differential equation. To maintain consistency between It\={o} and Stratonovich models, the drift term must be modified to ensure that Boltzmann equilibrium is obtained at long times in the absence of non-conservative forces (in our case, the applied current). The fundamental reason is that the SDE is simply a model of the underlying dynamics subject to two constraints: the chosen form of the thermal noise and the steady-state equilbrium Boltzmann distribution~\cite{Lau,Pesce}. In the absence of applied currents, (\ref{eq:eevolve}) can be written more concisely as:

\be
\langle\partial_{\mathrm{t}}\epsilon\rangle=\left[-\alpha f(\epsilon)+h(\epsilon)\right]+\sqrt{\frac{2\alpha}{\xi}f(\epsilon)}\cdot\dot{W}
\ee
with
\bea
f(\epsilon)=2\left[D(D+1)\langle m_z^2\rangle+\epsilon(1+\epsilon)\right],
\eea
where $h(\epsilon)$ represents the extra modification necessary in the drift term to retain all physically relevant Boltzmann relaxation properties. Deriving the It\={o} Fokker-Planck equation relative to such a dynamic then gives:

\be
\partial_t\rho=\partial_{\epsilon}\left[(\alpha f(\epsilon)-h(\epsilon)+\frac{\alpha}{\xi}\partial_{\epsilon}f(\epsilon))\rho+\frac{\alpha}{\xi}f(\epsilon)\partial_{\epsilon}\rho\right].
\ee
Upon imposing $h(\epsilon)\equiv\frac{\alpha}{\xi}\partial_{\epsilon}f(\epsilon)$, the steady-state solution reduces to the simple form $\rho_{\mathrm{eq}}(\epsilon)\propto\exp[-\xi\,\epsilon]$ as expected.

Employing the previously derived expression for $\langle m_z^2\rangle$ from Appendix A, $h(\epsilon)$ is found to be (in terms of the auxiliary variable $\gamma$):
\bea
h(\epsilon)&=&\frac{\alpha}{\xi}\frac{D(1-\gamma^2)+1}{1-\gamma^2}\left[1-\left(\frac{D(1-\gamma^2)+2}{D(1-\gamma^2)+1}\right)\frac{\mathrm{E}[1-\gamma^2]}{\mathrm{K}[1-\gamma^2]}+\frac{1}{\gamma^2(2-\gamma^2)}\left(\frac{\mathrm{E}[1-\gamma^2]}{\mathrm{K}[1-\gamma^2]}\right)^2\right]\nonumber\\
&+&\frac{\alpha}{\xi}\frac{D(1+\gamma^2)+1}{D(1-\gamma^2)+1},
\eea
which can be shown to lead to a negligible correction of the drift dynamics ($\approx 0.1\,\alpha/\xi\approx 10^{-5}$ since typical parameter values are $\alpha\sim0.01$ and $\xi\sim 100$).

\end{document}